\documentclass[]{article}
\usepackage{latexsym}
\usepackage{epsfig}

\Large

\long\def\@makecaption#1#2{%
  \vskip\abovecaptionskip
  \sbox\@tempboxa{\pushziti\small\rm\songti\zihao{-5}#1: #2\popziti}%
  \ifdim \wd\@tempboxa >\hsize
    \box\@tempboxa\par
  \else
    \global \@minipagefalse
    \hbox to\hsize{\hfil\box\@tempboxa\hfil}%
  \fi
  \vskip\belowcaptionskip}

\begin{document}

\centerline{\Large{\bf Some Important Concepts in Nonstandard}}
\centerline{\Large{\bf Analysis Theory of Turbulence (The
revised)}}

$${}$$
\centerline{Feng \quad Wu }\centerline{\footnote{klfjgjg}}
\centerline{\it Department of Mechanics
and Mechanical Engineering,}

\centerline{\it University of Science and Technology of China, Hefei 230026, China}
$${}$$
\setlength{\baselineskip}{25pt}

\noindent \small{Some important concepts in the nonstandard
analysis theory of turbulence are presented in this article. The
structure of point, on which differential equations are defined,
is analyzed. The distinction between the uniform point and the
non-uniform point, as well as between the standard point and the
nonstandard point, is showed. A new kind of equations, which
differ essentially from those in existent theory, is emphasized.
These new equations can hold at non-uniform points. The
applicability of the Navier-Stokes equations to turbulence is
discussed. Some illustrations of the nonstandard analysis theory
of turbulence are given too.}
$${}$$
\noindent PACS \quad 47.27.Ak

$${}$$

\large \setlength{\baselineskip}{25pt}
\section{Introduction}\indent
\indent A new approach, the nonstandard analysis picture, to the
theory of turbulence was presented in the paper \cite{lgc}. The
nonstandard analysis theory of turbulence(NATT) is based on the
nonstandard analysis mathematics and the fact that hierarchical
structure is universally existent in the world. The theory shows
that in a laminar flow, a particle of fluid is taken as uniform
wholly, and there does not exist any structure in the particle; on
the other hand, a particle of fluid in a turbulent field should
not be uniform wholly and some interior structure occurs.

\indent There are six assumptions in the new theory of turbulence.
They are:

\begin{quote}
    \it{Assumption 1: Global turbulent field is composed of standard
   points, and every standard point is yet a monad. Each monad possesses
   the internal structure, namely a monad is also composed of infinite
   nonstandard points (so called interior points of the monad)}.
\end{quote}
\begin{quote}
     \it{Assumption 2: The flows in monad fields are controlled by the
      Navier-Stokes equations.}.
\end{quote}
\begin{quote}
     \it{Assumption 3: Turbulent field is continuous}.
\end{quote}
\begin{quote}
     \it{Assumption 4: When a measurement at any point (monad) $(x_{1},x_{2},x_{3},t)$
     in a physical field is taken, the operation of the measurement
    will act randomly on one interior point (nonstandard point) of the point $(x_{1},x_{2},
    x_{3},t)$}.
\end{quote}
\begin{quote}
     \it{Assumption 5: When a measurement at any point (monad) of a turbulent
    field is made, the operation of the measurement will act in equiprobability on various
    interior points of the monad. This assumption is called the equiprobability
    assumption}.
\end{quote}
\begin{quote}
  \it{Assumption 6: In both the value and structure
  of function, physical function, defined on the interior points of the monads
  of a turbulent field, is infinitely close between two monads,  when these two monads
  are infinitely close to each other}.
\end{quote}

\indent Based on these assumptions, it is presented that the real
turbulent fluctuation stems from the uncertainty of measurement of
turbulence. The average of physical quantities is taken over the
point(monad), namely the point(monad)-average is adopted and
computed. And the fundamental equations of turbulence are obtained
too. The closure problem is overcome easily. The closed equations
are:

\noindent Choice one:
\begin{equation}
\frac{\partial \widetilde{U_{i}}}{\partial x_{i}}=0,\quad
\frac{\partial\widetilde{U_{i}}}{\partial t}+\frac{\partial
\widetilde{U_{i}}\widetilde{U_{j}}}{\partial
x_{j}}=-\frac{1}{\rho}\frac{\partial \widetilde{P}}{\partial
x_{i}}+\nu \nabla^{2}\widetilde{U_{i}}+0(\varepsilon^{2})
\end{equation}

\noindent Choice two:
\begin{equation}
\frac{\partial U_{i}}{\partial x_{i}}=0,\quad \frac{\partial U_{i}
}{\partial t}+\frac{\partial U_{i}U_{j}}{\partial
x_{j}}=-\frac{1}{\rho}\frac{\partial P}{\partial
x_{i}}+\nu\nabla^{2}U_{i}
\end{equation}
\begin{equation}
\frac{\partial u_{i}}{\partial x_{i}}=0, \frac{\partial
u_{i}}{\partial t}+U_{j}\frac{\partial u_{i}}{\partial
x_{j}}+u_{j}\frac{\partial U_{i}}{\partial x_{j}}-2\frac{\partial
u_{i}u_{j}}{\partial x_{j}}=-\frac{1}{\rho}\frac{\partial
p}{\partial x_{i}}+\nu\nabla^{2}u_{i}+0(\varepsilon^{3})
\end{equation}

\noindent Choice three:
\begin{equation}
\frac{\partial \widetilde{U_{i}}}{\partial x_{i}}=0,\quad
\frac{\partial\widetilde{U_{i}}}{\partial t}+\frac{\partial
\widetilde{U_{i}}\widetilde{U_{j}}}{\partial x_{j}}+\frac{\partial
u_{i}u_{j}}{\partial x_{j}}=-\frac{1}{\rho}\frac{\partial
\widetilde{P}}{\partial x_{i}}+\nu
\nabla^{2}\widetilde{U_{i}}+0(\varepsilon^{3})
\end{equation}
\begin{equation}
\frac{\partial u_{i}}{\partial x_{i}}=0,\quad \frac{\partial
u_{i}}{\partial t}+\widetilde{U_{j}}\frac{\partial u_{i}}{\partial
x_{j}}+u_{j}\frac{\partial\widetilde{U_{i}} }{\partial
x_{j}}=-\frac{1}{\rho}\frac{\partial p}{\partial
x_{i}}+\nu\nabla^{2}u_{i}+0(\varepsilon^{3})
\end{equation}

\indent Here $\varepsilon$ is the linear dimension of monad, yet
nonstandard number infinitesimal. And
$$U_{i}=\widetilde{U_{i}}+u_{i},\quad P=\widetilde{P}+p,$$
$``\sim"$ expresses the average over monad. $U_{i}$ and $P$ are
the instantaneous quantities of velocity and pressure
respectively.

\indent After further contemplation over the NATT, we can present
some important concepts. These concepts are discussed as follows.

\section{On the concept of ``point"}\indent

\indent When physical phenomena (e.g., complex phenomena) are
studied, people's attention is usually concentrated on inquiry
into the characteristic of the equations, governing the phenomena,
and their solutions. Is it possible that the phenomena are
discussed from other angle? It is well known that differential
equations are always defined on and applied to points. A point in
mathematics is absolute geometric point; but a physical point, in
fact, is micro-volume. And a physical point, in some cases, has
the structure in it. Therefore, we could observe and comprehend
physical phenomena from the angle of analyzing the nature of the
point, to which the governing equations of the phenomena are
applied.

\indent Point is an abstract concept of mathematics, surely the
concept of point is drawn from objective physical reality. In
fluid mechanics, for example, a particle of fluid is taken, in the
abstract, as a point. In a laminar flow, a point(a particle of
fluid) is uniform and no structure exists through the particle. On
the other hand, in turbulence, a point (a particle of fluid)
possesses interior structure and is non-uniform. A particle of
fluid in the two cases, laminar flow and turbulence, is abstracted
as a point. However, the former is uniform point and the latter
non-uniform point. Every point(particle of fluid), which has
interior structure under some conditions, is formed of numerous
fluid-particles in lower level. A fluid-particle in lower level
can be thought of as uniform and abstracted as a uniform point.
Yet every point(particle of fluid) of a flow field is called as
monad; and a fluid-particle in lower level is called as an
interior point of the monad in paper \cite{lgc}.

\indent Moreover, a particle of fluid is abstracted as a standard
point and a fluid-particle in lower level is abstracted as a
nonstandard point in paper \cite{lgc}. A standard point(monad)
corresponds to a real number, and a nonstandard point to a
nonstandard number. Here we give the other meaning of standard
points. The dimension of the standard point is an infinitesimal in
the level, of which the characteristic dimension could compare
with the dimension of human being self. In that level, people do
many practical activities, such as navigation, aviation,
spaceflight etc.. The standard point is proper for these practical
activities, i.e., the point in the physical models related to
these activities corresponds just to the standard point. The
dimension of standard points is not determined by people at will,
but by the nature of physical laws and the characteristic of the
practical activities. Though we can not show exactly how large a
standard point is, the dimension of standard points is objective.
Similarly, the dimension of the nonstandard point is an
infinitesimal in lower level.

\indent The meaning of ``point", in a word, is not fixed and
absolute. By the concepts of nonstandard analysis, ``point"
possesses plentiful and vital content. There is need to make a
distinction between different points, i.e., the uniform point and
the non-uniform point, the standard point and the nonstandard
point.

\section{Two kinds of differential equations}\indent

\indent Physical laws usually are expressed by differential
equations in mathematics. Yet the differential terms in the
equations of existent theory are as follows:
 \begin{equation}
\frac{\partial f}{\partial t}=\lim_{\triangle t\rightarrow
0}\frac{f(t+\triangle t)-f(t)}{\triangle t},\quad \frac{\partial
f}{\partial x}=\lim_{\triangle x\rightarrow 0}\frac{f(x+\triangle
x)-f(x)}{\triangle x}
\end{equation}
\noindent Therefore, the limit in mathematics means the fundament
of the differential equations in existent theory, namely these
equations are in the frame of $\delta-\varepsilon$. The
limits$(\triangle t\rightarrow 0,\triangle x\rightarrow 0)$ in
these equations mean that $\triangle t$ and $\triangle x$ tend to
absolute zero in mathematics, while tend to uniform point(uniform
particle) in physics. This fact sets a limit to the nature of the
points, to which these equations are applicable. These equations
are applicable only to uniform points, rather than non-uniform
points.

\indent On the other hand, the fundamental equations of turbulence
were presented in the paper \cite{lgc}. The equations have the
same form as those in existent theory. But the ``differential"
terms in the equations of the new theory(NATT) are:
\begin{equation}
\frac{\partial f}{\partial
t}=\frac{f(t+\varepsilon_{t})-f(t)}{\varepsilon_{t}},\quad
\frac{\partial f}{\partial
x}=\frac{f(x+\varepsilon_{x})-f(x)}{\varepsilon_{x}}
\end{equation}

\noindent Here $\varepsilon_{t}$ and $\varepsilon_{x}$ are the
infinitesimals, which are the linear dimensions of monads, rather
than arbitrary infinitesimals. In nonstandard analysis,
$\varepsilon_{t}$ and $\varepsilon_{x}$ are certain
numbers(nonstandard numbers). So there is no limit term in these
equations. Obviously, these ``differential equations" are not
based on the limits$(\triangle t\rightarrow 0, \triangle
x\rightarrow 0)$ and are out of the frame of $\delta-\varepsilon$.
In fact, a new kind of equations was presented in paper
\cite{lgc}. The new equations differ essentially from those in
existent theory. Here the term in new equations, by convention, is
still written as the differential in form. But the meaning of the
term is different from that of ordinary differential. New kind of
equations can hold at non-uniform points. Obviously, these new
equations have more natural relation with discretization form of
equations than those based on the frame of $\delta-\varepsilon$ in
numerical computation.

\indent Therefore, there are two kinds of equations: One is the
equation in existent theory. This equation is based on
limits$(\triangle t\rightarrow 0, \triangle x\rightarrow 0)$(or so
called in the frame of $\delta-\varepsilon$) and applicable only
to uniform points. The other is the equation in the NATT. The
second kind of equations is out of the frame of
$\delta-\varepsilon$, and based on the nonstandard analysis. These
new equations can hold at non-uniform points. When motion of fluid
is slow and varies small, the description of the motion by the
first kind of equations is suitable. Otherwise, when the motion of
fluid is very fast and varies drastically, e.g. in the case of
turbulence, the reasonable equations describing the motion are not
the first kind of equations, but the second kind of equations.
Surely, the second kind of equations becomes the first kind of
equations, i.e. there is no difference between the two kinds of
equations at the uniform point.

\section{Applicability of the Navier-Stokes Equations}\indent

\indent Surely to answer whether the Navier-Stokes equations are
applicable to the turbulence is important. Now this question is
presented by following statements: ``The Navier-Stokes equations
hold in laminar flows.", ``The Navier-Stokes equations hold in
turbulence too.", ``Do the Navier-Stokes equations hold in
turbulence?" etc.. Obviously, these statements show that when
thinking of this question, people always pay their attention to
that a flow is laminar or turbulence. Usually the Navier-Stokes
equations are thought of as applicable to laminar flows. And some
think that the Navier-Stokes equations hold in turbulence, others
do not.

\noindent The Navier-Stokes equations:
\begin{equation}
\frac{\partial U_{i}}{\partial x_{i}}=0,\quad \frac{\partial U_{i}
}{\partial t}+\frac{\partial U_{i}U_{j}}{\partial
x_{j}}=-\frac{1}{\rho}\frac{\partial P}{\partial
x_{i}}+\nu\nabla^{2}U_{i}
\end{equation}

\indent It is well known that the Navier-Stokes equations are
based on the limits$(\triangle t\rightarrow 0,\triangle
x\rightarrow 0)$. Therefore, the Navier-Stokes equations hold only
at uniform points. In laminar flows, not only nonstandard but also
standard points are uniform. So the Navier-Stokes equations hold
in laminar flows. However, in turbulence, nonstandard points are
uniform, while standard points are not uniform and possess
interior structure. Hence, in turbulence the Navier-Stokes
equations hold only at nonstandard points, rather than at standard
points. Therefore, the key of the applicability of the
Navier-Stokes equations lies in that the Navier-Stokes equations
do not hold at non-uniform points, but only uniform points.

\indent Therefore, if the turbulent field is thought as composed
of nonstandard points, i.e., the turbulent field is thought as in
one level only rather than possesses hierarchical structure, the
Navier-Stokes equations hold in the turbulence. But this viewpoint
has only theoretical and abstract meanings. In engineering
practice, the points, corresponding to the engineering practice,
are the standard points(monads). The Navier-Stokes equations do
not hold on these standard points(monads), i.e., the Navier-Stokes
equations are not applicable to turbulence. In other word, the
Navier-Stokes equations are applicable to monad field rather than
the global turbulent field.

\section{Some Illustrations of the NATT}\indent

\indent The equations (1)-(5) are the new kind of equations, which
can hold on the non-uniform points. For example, they can hold on
the standard points(monads) in turbulence. But the Navier-Stokes
equations hold only on the uniform points. In the case of
turbulence, the Navier-Stokes equations hold only on the
nonstandard points. It is obvious logically that the
discretization of the Navier-Stokes equations must be based on the
nonstandard points, while discretization for the equations (1)-(5)
are based on the standard points(monads) in numerical computation
of turbulence. The standard and nonstandard points are points in
different levels, the scale of standard point is $\sim
\varepsilon$(Here $\varepsilon$ is the scale of the monad of
global turbulent field.), but the nonstandard point has the scale
$\sim \varepsilon^{3}$. This is the reason why the Navier-Stokes
equations must be computed by use of very small grids, while the
equations(1)-(5) could be simulated by coarse grids in the case of
turbulence.

\indent By the analysis above, we know that when numerical
computation is taken for the Navier-Stokes equations, there must
be a limit to the grids obtained from discretization of the
Navier-Stokes equations. The limit is just that the grids must be
uniform, i.e. there is not structure through every grid wholly.
The direct numerical simulation(DNS) is generally recognized as a
good method of computation for the Navier-Stokes equations in
turbulence now. But the grid of discretization in DNS is not so
small that there is not structure in the grid.  It is known that
there still exists structure(so called smallest vortices) in the
discretization-grid of DNS. Therefore, the method of DNS is not
reasonable from the angle of the Navier-Stokes equations in
mathematics. But the results of the computation in DNS are very
well, why? The reason lies in that when DNS is made, people do not
really compute the Navier-Stokes equations, but those in the
NATT(i.e., the equations (1) in Choice one). The results of DNS
are not the values of instantaneous quantities at the standard
point, but average quantities over the grid. Hence, the success of
DNS could be taken for example of the reasonableness of the NATT.

\indent Moreover, it is well known that many researchers have used
coarse grids for turbulence
simulation\cite{kuw1}\cite{kuw2}\cite{dep}\cite{fle}. They
computed the Navier-Stokes equations by coarse grids in
turbulence. Their computations are not reasonable from the angle
of the Navier-Stokes equations. But the results of their
computations are well agreeable to the measurements. Also this is
an example for the reasonableness of the NATT, because what they
have simulated are not the Navier-Stokes equations but the
equations (1). The equations (1) are the new kind of equations,
which can hold on the standard points(monads). The coarse grids
are obtained from the discretization based on the standard points.

  $${}$$

\end{document}